\documentclass[11pt]{article}
\usepackage{mymoriond,times,epsf,subfigure,epsfig}

\newcommand{\model}{SU(2)$_{\rm L}\times$SU(2)$_{\rm R}\times$U(1)}
\newcommand{\Kd}{$K^0$--$\bar{K^0}$}
\newcommand{\Bd}{$B^0_d$--$\bar{B^0_d}$}
\newcommand{\Bs}{$B^0_s$--$\bar{B^0_s}$}
\newcommand{\ee}{$\epsilon'/\epsilon$}
\newcommand{\goldd}{$B_d\to J/\psi K_S$}
\newcommand{\golds}{$B_s\to J/\psi \phi$}

\begin{document}

%%%%%%%%%%% Titlepage
\vspace*{4cm}
\title{SPONTANEOUS CP VIOLATION IN THE LEFT--RIGHT--SYMMETRIC MODEL}

\author{Patricia Ball}

\address{CERN, Theory Division, 1211 Geneva 23, Switzerland}

\maketitle\abstracts{
We investigate the pattern of CP violation in \Kd, \Bd\ and \Bs\ mixing
in a symmetrical \model\ model with spontaneous CP
violation. Performing a careful analysis of all relevant restrictions
on the model's parameters from $\Delta M_K$, $\Delta M_{B_d}$, $\Delta
M_{B_s}$, $\epsilon_K$, the sign of \ee\  and the mixing-induced 
CP asymmetry in \goldd, we find that the mass of the right-handed 
charged gauge-boson, $M_2$, is restricted to be in the range 2.75 to 
13$\,$TeV, and that the mass of the flavour-changing neutral-Higgs 
bosons, $M_H$, must be between 10.2 to 14.6$\,$TeV. We also find that
the model, although still compatible with present experimental data,
cannot accomodate the SM prediction of large CP violation in 
\goldd, but, on the other hand, predicts a large CP-asymmetry of 
${\cal O}(40\%)$ in \golds. These specific predictions make it
possible to submit the model to a scrupulous test at B factories and
hadron colliders.\\[2cm]
{\it\large Invited Talk given at XXXVth Rencontres de Moriond,
  Electroweak Interactions and Unified Theories, Les Arcs, France,
  March 2000}}

\newpage

\section{Introduction}

In this talk I report on a recent investigation \cite{my,robi} 
of a non-standard mechanism of 
CP violation in an attractive extension of the Standard Model (SM),
the spontaneously broken left--right--symmetric model.\cite{complaints} 
Available data on the mass differences and the measured amount of CP violation
in the K and B system seriously constrain the model's parameter space.
Even if current theoretical uncertainties persist in the
K system, the expected experimental progress in B physics \cite{wrup} will soon
bring conclusive tests of the model in the form of precise values of
the CP violating asymmetries in the decays $B_d\to J/\psi K_S$ and
$B_s\to J/\psi \phi$, which we predict to sizeably deviate from their
SM expectations.

It is well known that CP is a natural symmetry of pure gauge theories
with massless fermions. Consequently, CP violation actually probes the
least known sector of unified theories, namely the scalar and Yukawa couplings.
The current development of dedicated accelerators to probe CP violation in
the B system prompts studies of possible departures from the
SM. Left--right (LR) symmetric models based on the extended gauge group
\model\ offer the 
advantage of a well-defined, and actually quite constraining context, 
largely testable experimentally, while presenting a structure
significantly different from the SM. 
Whereas in general LR models CP violation can come from several
sources, we consider here a restricted model which exhibits the
aesthetically attractive
feature of {\it spontaneous} CP violation. This means that the Lagrangian
exhibits manifest CP symmetry, which at low energies is broken by 
"misaligned" phases of the symmetry-breaking vacuum expectation value
(VEV). In this model,
spontaneous breakdown is the {\it only} source of CP violation. Under this
hypothesis, Ecker and Grimus have shown \cite{EGNPB} that, except for an
exceptional case (which will not be considered here), the Yukawa
couplings can be parametrized in terms of two real symmetrical matrices.
As a consequence, all CP-violating phases of the
model can be  be calculated exactly, as they are related to a unique
phase (denoted $\alpha$ below) which affects 
the SU(2)$_{\rm L}\times\,$SU(2)$_{\rm R}$ breaking VEV. 
This point is important, as it links
baryogenesis, and in particular the sign of the matter--antimatter
asymmetry, to low-energy CP violation.\cite{houart}

In practice, the above-defined "Spontaneously Broken Left--Right model"
(SB--LR) adds a very economical {\it four} parameters to the SM:
2 boson masses and 2 parameters describing the VEV that breaks
SU(2)$_{\rm L}\times\,$SU(2)$_{\rm R}$. Despite being an {\it extension} of the
SM, the SB--LR is in some sense {\it more restrictive}
than the SM itself. Indeed, while the SM is a subset of general LR obtained by 
sending the R-sector masses to infinity, a similar procedure applied
to the SB--LR yields additional constraints since the CKM phase $\delta$ is no
longer independent, but predicted within the model. To be specific, we
find that in the SB--LR $\delta$ is too small, 
$|\delta|<0.25$ or $|\delta-\pi|<0.25$,  whereas a recent global fit
\cite{fit} yields $\delta= 1.0 \pm 0.2$. Hence the SM limit of the
SB--LR is inconsistent by 3.5$\sigma$ with current experiments.
This has the important consequence that the SB--LR is actually testable,
and distinct from the SM: experimental bounds cannot be indefinitely evaded
by simply sending the R-sector to infinite masses: scalars and vectors in the
range (2--20) TeV are definitely needed.

Experimental constraints on the SB--LR, mainly from the K system, have
been thoroughly investigated in the late 80's.\cite{langacker} Since
then, many SM parameters, in particular the CKM angles and the top
quark mass, have been measured much more accurately, and also
theory has progressed, as exact relations for the
CP-violating phases in the quark mixing-matrices have been
derived,\cite{JMF} which supersede the previously used small-phase
approximation \cite{SPA} that breaks down for $b$ decays due
to the large top-quark mass. The
perspective of finding non-standard CP violation in the B system at
the B factories, the Tevatron and the LHC has prompted us to undertake
a new comprehensive analysis of restrictions on the SB--LR from
available experimental data. The 
main results 
can be summarized as follows:
\begin{itemize}
\item the role of the Higgs bosons, neglected in most analyses, is crucial;
\item the decoupling limit of the model, where the extra boson masses
  $M_2$ and $M_H$ are sent to infinity, is
  experimentally excluded, which implies {\em upper bounds} on $M_2$
  and $M_H$;
\item neglecting uncertainties of quark masses and CKM angles,
  the SB--LR favours opposite signs of the CP-violating observables
  ${\rm Re}\,\epsilon$ and $a_{\rm CP}(B_d\to J/\psi K_S)$, which are both
  expected to be positive in the SM; hence, the model cannot accommodate both
  the experimentally measured $\epsilon$ and the SM expectation
  $a^{\rm SM}_{\rm CP}(B_d\to J/\psi K_S)\approx 0.75$ and is
  excluded if $a_{\rm CP}$ will be measured close to its SM expectation;
\item the CP asymmetry in $B_s\to J/\psi \phi$, which is negligible in
  the SM, can be as large as 35\% in SB--LR.
\end{itemize}

\section{The Left-Right-Symmetric Model with Spontaneous CP 
Violation}\label{sec:LR-Model}

Before discussing its predictions for CP-violating phenomena, let us
explain very shortly the essential features of the SB--LR. As
already mentioned, it
is based on the gauge group \model, 
which cascades down to the unbroken electromagnetic subgroup
U(1)$_{\rm em}$ through the following simple symmetry-breaking
pattern:
$$
\underbrace{\mbox{SU(2)}_{\rm R}\times \mbox{SU(2)}_{\rm L}}_{
\underbrace{\displaystyle \mbox{SU(2)}_{\rm L}\hspace*{0.6cm}
  \times \mbox{U(1)}}_{\displaystyle \mbox{U(1)}_{\rm em}}
\hspace*{-2.4cm}}\times \mbox{U(1)}
$$
The scalar sector is highly model-dependent; for the
generation of quark masses, there has to be
 at least one scalar bidoublet $\Phi$, i.e.\ a
doublet under both SU(2), which, by spontaneous breakdown of
SU(2)$_{\rm R}\times\,$SU(2)$_{\rm L}$, acquires the VEV
\begin{equation}\label{eq:VEV}
\langle \Phi \rangle = \frac{1}{\sqrt{2}}\left( \begin{array}{cc} v &
    0\\ 0 & w e^{i\alpha}\end{array}\right).
\end{equation}
Here,  $v$ and $w$ are real and the phase $\alpha$  is the (only) source of
CP violation in the model. The particle content of $\Phi$ corresponds to 
four particles, one analogue of the SM Higgs, two 
flavour-changing neutral Higgs bosons, and one flavour-changing charged 
Higgs. The masses of these new Higgs particles can be assumed to be 
degenerate to good accuracy; they will be denoted by $M_H$ below.

LR symmetry implies that the left-handed quark sector of the SM 
gets complemented by a right-handed one, with quark mixing matrices 
$V_{\rm L}$ and $V_{\rm R}$, respectively, and $|V_{\rm L}|=|V_{\rm
  R}|$ (but $V_{\rm L}\neq V_{\rm R}$ due to different complex phases!).
 In the standard Maiani convention, 
$V_{\rm L}$ contains one, $V_{\rm R}$ five complex phases, which depend on 
the three generalized Cabibbo-type angles (``CKM angles''), the quark masses, 
and the VEV (\ref{eq:VEV}). The presence of such a large number of weak 
phases, calculable in terms of only one non-SM
variable,\footnote{I.e.\ the
  variable $\beta$ introduced in the next section; note also that
  there is a 64-fold discrete ambiguity of the phases due to the {\it
    signs} of the quark masses, which are physical in LR models.} is a
feature that 
makes the investigation of CP-violating phenomena in the SB--LR very 
interesting. The left- and right-handed charged gauge-bosons $W_{\rm L}$ and 
$W_{\rm R}$ mix with each other;  the mass eigenstates are denoted by
$W_1$ and $W_2$. The mixing angle 
$\zeta$, obtained as
\begin{equation}
\zeta = \frac{2 |vw|}{|v|^2 + |w|^2}\left( \frac{M_1}{M_2}\right)^2,
\end{equation}
is rather small: as the 
ratio $|v|/|w|$ is smaller than 1,\footnote{Which can always be
achieved by a redefinition of the Higgs bidoublet $\Phi\to \sigma_2
\Phi^* \sigma_2$.} one has $\zeta < (M_1/M_2)^2 \sim 10^{-3}$
(assuming $M_2$ in the TeV range as indicated by the experimental
absence of right-handed weak currents).
There are, however, arguments according to which a small ratio
$|v|/|w|\sim {\cal O}(m_b/m_t)$ would naturally explain the observed
smallness of the CKM angles; \cite{natural} in this case, $\zeta \sim
10^{-5}$. An experimental bound on $\zeta$ 
can in principle be obtained from the
upper bound on the electromagnetic dipole moment of the neutron, which
is induced by L--R mixing; existing theoretical calculations are,
however, very sensitive to the precise values of only poorly known
nucleon matrix elements; 
\cite{had}\footnote{In addition, the Higgs contributions
  to the dipole moment are usually not included.}
 the present status of an
experimental bound on $\zeta$ is thus not quite clear, although large values
of $\zeta \sim 10^{-4}$ appear to be disfavoured.

The fact that the SB--LR has no perceptible impact on SM
tree-level amplitudes (no experimental indication of right-handed weak
interactions or large flavour-changing neutral currents!) implies that
the new boson masses must be in the TeV range. The SB--LR effects thus
manifest themselves mostly in
\begin{itemize}
\item $W_{\rm L}$--$W_{\rm R}$ mixing in top-dominated penguin diagrams, 
enhanced by large quark-mass terms from spin-flips, $\zeta\to \zeta\,
  m_t/m_b$ (similar for penguins with charged-Higgs particles);
\item SM amplitudes that are forbidden or heavily suppressed 
  (e.g.\ electromagnetic
  dipole moment of the neutron);
\item mixing of neutral K and B mesons, where the suppression
  factor $(M_1/M_2)^2$ is partially compensated by large
  Wilson-coefficients or hadronic matrix-elements (chiral enhancement
  in K mixing), and to which the flavour-changing Higgs bosons contribute 
  at tree level. 
\end{itemize}
The SB--LR does, however, {\it not} significantly modify the decay amplitudes
of the ``gold-plated'' decay mode $B_d\to
J/\psi K_S$ or the decay $B_s\to J/\psi\phi$:$\,$\footnote{Recall that, in the
SM and using the standard ``generalized Wolfenstein parametrization''
of the CKM matrix, the amplitudes of these decays carry only small
or zero weak phases
and the CP-violating asymmetry is essentially given by the
B--$\overline{\mbox{B}}$ mixing-phase.} these
 decays are dominated by one single CKM amplitude ($b\to
ccs$) with contributions from colour-suppressed tree and
penguin-topologies with internal $c$ or $t$ quarks. The
``gold-plated'' mode \goldd\ is the standard example for a special
type of decays of a
neutral B meson into a CP-eigenstate, whose time-dependent
CP-asymmetry takes a particularly simple form and allows the direct
extraction of a weak CKM phase without hadronic uncertainties:
\begin{equation}\label{golden_plates_or_what?}
{\cal A}_{\rm CP}(B_d\to J/\psi K_S) = 
\frac{\Gamma(t)-\overline{\Gamma}(t)}{\Gamma(t)+\overline{\Gamma}(t)}=
\sin\phi_{\rm weak}\, \sin \Delta M_d t,
\end{equation}
where $\overline{\Gamma}(t)$ denotes the decay-rate of $\bar
B^0_d(t)\to J/\psi K_S$ and $\Delta M_d$ is the mass-difference in the 
$B^0_d$--$\bar B^0_d$ system. The LR contribution to the tree-topology 
is given by $W_L$--$W_R$ mixing, $W_R$ or 
neutral-Higgs exchange and suppressed by $\sim(M_1/M_2)^2\sim 10^{-3}$ or more
with respect to the SM contribution. As for the penguins, internal
$W_R$ or charged-Higgs exchange are suppressed by the same order of
magnitude as for the tree-topology, and the only potentially relevant 
contribution comes from $W_L$--$W_R$ mixing: the corresponding
top-penguin topology is enhanced by a spin-flip factor $\sim \zeta m_t/m_b$,
which is at most 5\% in the most unfavourable case $|v|/|w| = 1$ and in
the range of permille for the preferred value $|v|/|w|\sim {\cal O}(m_b/m_t)$.
Consequently, the SB--LR contributions to the amplitudes 
of the ``gold-plated'' B decays are small and
do neither scratch the lustre of the golden plates nor 
yield sizeable direct CP violation:$\,$\footnote{Which would show up as
  terms in $\cos \Delta M_d t$ in (\ref{golden_plates_or_what?}).} 
the main impact of the
model on these decays is from B--$\overline{\mbox{B}}$ mixing and
modifies the extracted value of the weak phase $\phi_{\rm
  weak}$. Sizeable LR-effects on amplitudes are, however, 
to be expected in theoretically less
``clean'' channels like $B_d\to \pi\pi$ and $b\to s\gamma$.

\section{Phenomenological Analysis}\label{sec:analysis}

As mentioned above, the main impact of the SB--LR on channels with
available experiemental data is to modify the SM pattern of neutral K
and B meson mixing. The relevant quantity to be calculated is the
matrix-element 
$$
\langle M^0 \mid {\cal H}_{\rm eff}^{|\Delta F|=2}\mid \bar M^0\rangle
= 2 m_M M_{12}
$$
with the effective weak Hamiltonian ${\cal H}_{\rm eff}^{|\Delta
  F|=2}$. Experimental observables which restrict SB--LR contributions
to $M_{12}^{B,K}$ are
\begin{equation}
\Delta M_{B,K} = 2 \left| M_{12}^{B,K}\right|,\quad
\epsilon_K\approx \frac{1}{2\sqrt{2}}\, e^{i\pi/4}\,\sin (\arg
M_{12}^K),\quad
a_{\rm CP}(B_d\to J/\psi K_S)  =  \sin (\arg
M_{12}^{B_d}).
\end{equation}
In addition, we also consider constraints posed by the smallness
of direct CP violation in the K system encoded in the value of
$\epsilon'/\epsilon$, but in view of the theoretical uncertainties
associated with this observable \cite{talk2} we only require the {\it
  sign} of $\epsilon'/\epsilon$ to be correctly reproduced. 
$|M_{12}^K|$ is affected by long-distance QCD uncertainties which are
also present in the SM, so that in our analysis, instead of attempting
a full calculation of that quantity, we impose the reasonable
requirement that SB--LR effects be smaller than the experimental
mass difference: $2 |M_{12}^{K,SB-LR}|< \Delta M_K$. The quantities 
$\arg M_{12}^K$
and $M_{12}^B$, on the other hand, are short-distance dominated and 
can be reliably calculated in the SB--LR. For the technicalities I
refer to Ref.~\cite{my}; here I only specify the set of SM input
parameters,\footnote{Note that we do not take into account
  uncertainties associated with these parameters.}
\begin{equation}\label{eq:input}
\renewcommand{\arraystretch}{1.4}
\begin{array}[b]{lcllcllcl}
\overline{m}_t(\overline{m}_t) & = & 170\,{\rm GeV}, &
\overline{m}_b(\overline{m}_b) & = & 4.25\,{\rm GeV},\\
\overline{m}_c(\overline{m}_c) & = & 1.33\,{\rm GeV}, & 
\overline{m}_s(2\,{\rm GeV}) & =& 110\,{\rm MeV},\\
m_s/m_d & = & 20.1, & m_u/m_d & = & 0.56,\\
\multicolumn{6}{l}{|V_{us}|  =  0.2219,\quad |V_{ub}|  =  0.004,\quad
|V_{cb}| = 0.04,}
\end{array}
\renewcommand{\arraystretch}{1}
\end{equation}
and remind that $M_{12}$ depends in addition on the SB--LR parameters
$M_2$, $M_H$ and, as shown in Ref.~\cite{JMF}, the variable $\beta$
defined as
\begin{equation}
\beta = \arctan \,\frac{2 |wv| \sin \alpha}{|v|^2-|w|^2}\,.
\end{equation}

The combined analysis of the experimental data on $\Delta
M_{K,B_d,B_s}$, $\epsilon_K$, the sign of $\epsilon'/\epsilon$ and
$a^{\rm exp}_{\rm CP}(B_d\to J/\psi K_S)=-0.79^{+0.44}_{-0.41}$ (see the talk
\cite{talk})
yields the allowed region for $M_2$ and $M_H$ shown in Fig.~1 and the
correlations between values of $a_{\rm CP}$ and $\epsilon_K$ shown in
Fig.~2.
\begin{figure}[tb]
\begin{minipage}[b]{0.48\textwidth}
\epsfxsize=\textwidth\epsffile{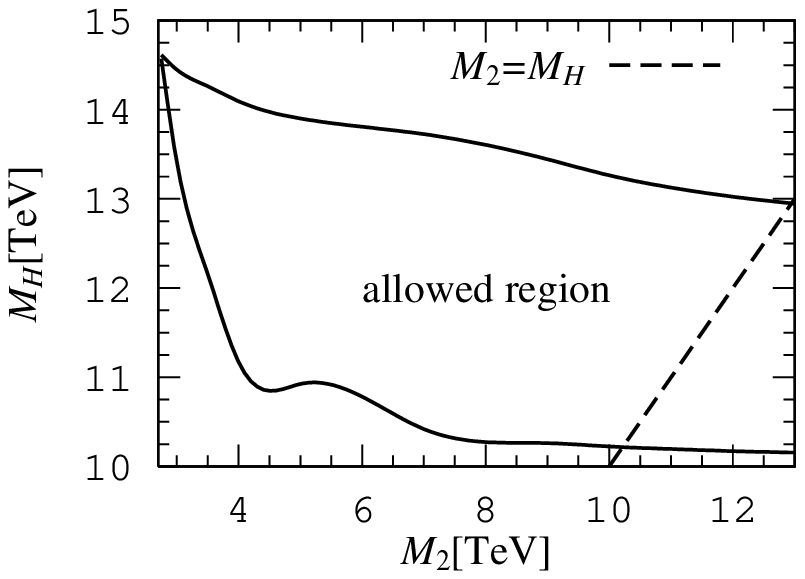}
%\vspace*{-0.5cm}
\caption[]{Allowed region in $(M_2,M_H)$.}
\end{minipage}
\hspace*{10pt}
\begin{minipage}[b]{0.48\textwidth}
\epsfxsize=\textwidth\epsffile{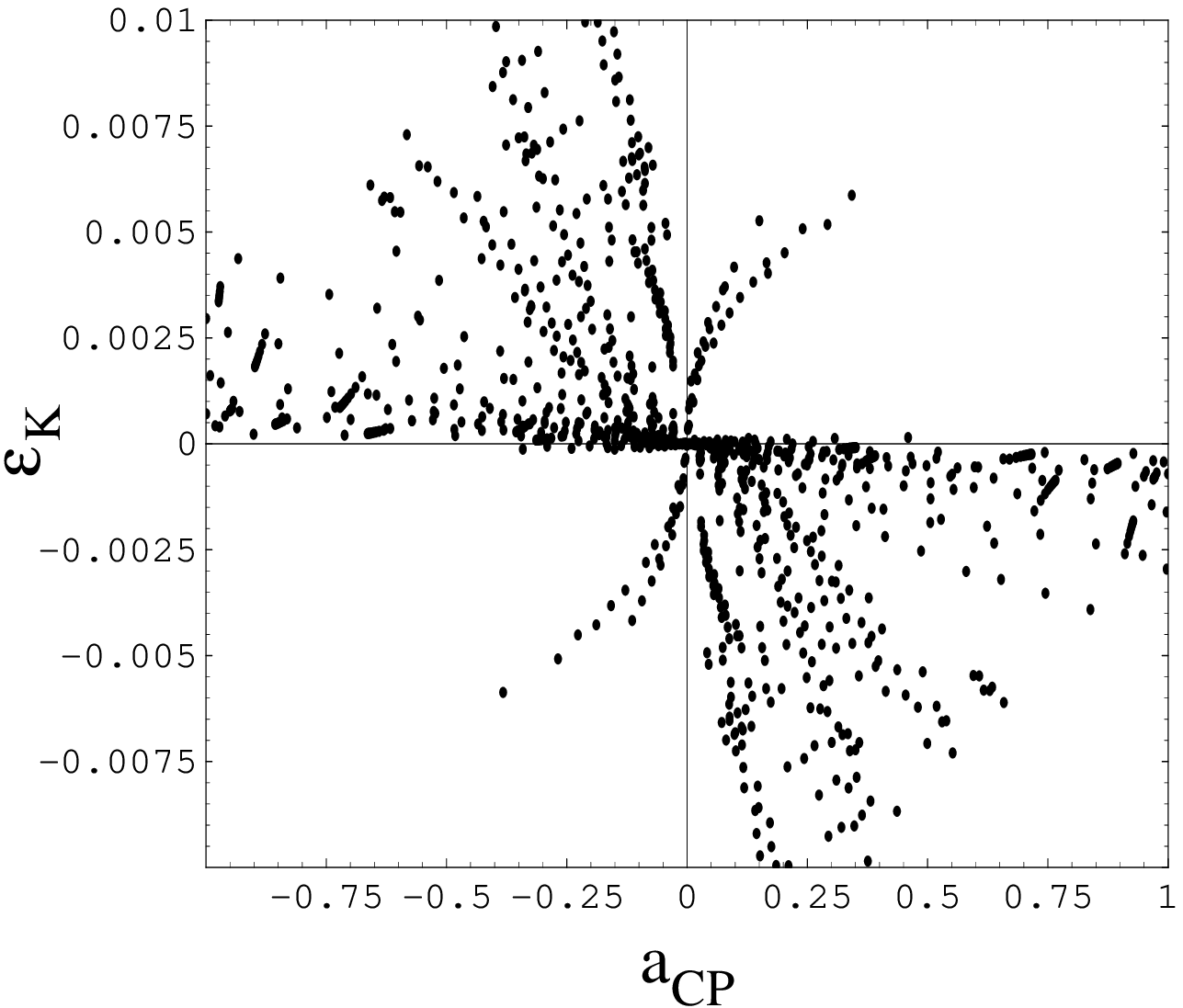}
%\vspace*{-0.5cm}
\caption[]{Allowed values of  $\epsilon_K$ and 
$a_{\rm CP}(B_d\to J/\psi K_S)$.}
\end{minipage}
\end{figure}
The evident preference of the SB--LR for {\it opposite} signs of
$a_{\rm CP}$ and $\epsilon_K$, which is in contradiction to experiment
at  98\% CL, \cite{talk} actually helps to resolve the 64-fold
discrete ambiguity of the CKM phases mentioned in the previous
section: only {\it one} of these 64 solutions can reproduce a
positive $\epsilon_K$ compatible with the experimental result {\it
  and} $a_{\rm CP}>0$. The resulting value of $a_{\rm CP}$ is,
however, rather smallish, $a_{\rm CP}<0.1$, and at variance with the SM
expectation $a_{\rm CP}^{\rm SM}\approx 0.75$, but in agreement with
the present experimental result within 2$\sigma$. I thus quote as a first
specific and testable prediction of the SB--LR:
$$
\fbox{$a_{\rm CP}^{\rm SB-LR}(B_d\to J/\psi K_S) < 0.1
  \Longleftrightarrow a_{\rm CP}^{\rm SM}(B_d\to J/\psi K_S) \approx
  0.75$}
$$

With the parameters $M_2$, $M_H$ and $\beta$ being constrained, we can
now predict the allowed range for mixing-induced CP violation in the
$B_s$ system. In the simple case of $b\to ccs$ dominated $B_s\to f$
transitions into a final state $f$ that is a CP eigenstate (e.g.\
$f=D_s^+ D_s^-$, $J/\psi \eta(')$), the CP asymmetry is completely
analogous to the $B_d\to J/\psi K_S$ case:
$$
a_{\rm CP}(B_s\to D_s^+ D_s^-, J/\psi \eta(')) = \sin (\arg
M_{12}^{B_s});
$$
the corresponding correlation with $a_{\rm CP}(B_d\to J/\psi K_S)$ is
plotted in Fig.~3(a). The situation is a bit more complicated for the
decay $B_s\to J/\psi\phi$, as here the final state is a superposition
of CP-even and odd states. The standard time-dependent CP asymmetry 
as defined in (3) does now contain poorly known hadronic
matrix-elements. These can, in
principle, be extracted by an analysis
of the angular correlations between the observed decay products
$(J/\psi\to)\ell^+\ell^-$ and $(\phi\to)K^+K^-$ \cite{amol}. Such an analysis
requires, however, large statistics and is probably possible only
at the LHC.\cite{wrup} 
\begin{figure}[tb]
\subfigure[Allowed values for $a_{\rm CP}
(B_d\to J/\psi K_{\rm S})$ and 
$a_{\rm CP}(B_s\to f)$, with 
$f=D_s^+D^-_s$, $J/\psi\, \eta^{(')}$.]
{\epsfig{file=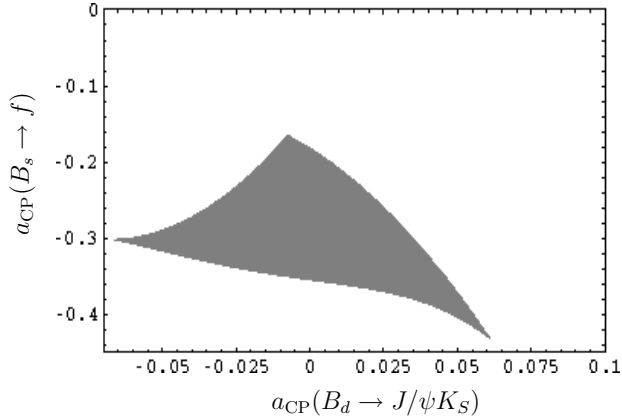,width=0.48\textwidth}}
\subfigure[Correlation between $\Delta M_s$ and $\Delta\Gamma_s$, 
normalized to their SM values.]
{\epsfig{file=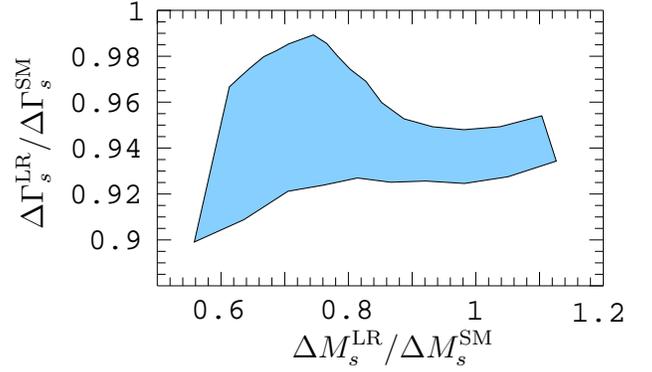,width=0.48\textwidth}}
\hspace*{10pt}
\subfigure[The time-dependent CP-asymmetry $a_{\rm CP}(B_s(t)\to
J/\psi\phi)$.]{\epsfig{file=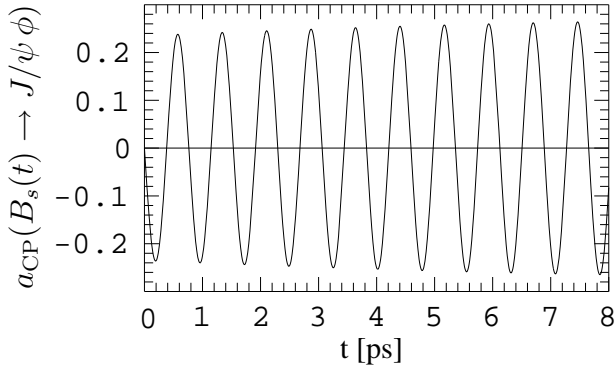,width=0.48\textwidth}}
\subfigure[$A_{\rm CP}(B_s\to J/\psi\phi)$ as a function
of the hadronic parameter $D$.]{\epsfig{file=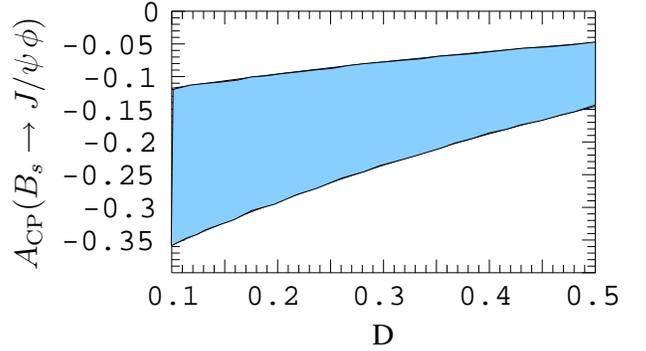,width=0.48\textwidth}}
\vspace*{-0.5cm}
\caption[]{Predictions of the left-right symmetric model for several
  CP observables in $B_s$ decays.}\label{fig:LRSM}
\end{figure}

In Fig.~3(b), we show the correlation 
between the $B^0_s$--$\overline{B^0_s}$ mass and width difference
$\Delta\Gamma_s$ in the 
SB--LR. The reduction of $\Delta\Gamma_s$ through new-physics effects
is not very effective 
in this case, whereas the mass difference $\Delta M_s$ may be reduced
significantly. Although, at first glance, values of $\Delta M_s$ as small 
as $0.55\Delta M_s^{\rm SM}$ may seem to be at variance with the 
experimental bound $\Delta M_s > 14.3\,{\rm ps}^{-1}$ at 95\% 
CL\ \cite{blaylock}, this is actually 
not the case: with the hadronic parameters from \cite{laurent} and 
$|V_{ts}| = 0.04$ with the generalized Cabibbo-angles fixed from 
(\ref{eq:input}), one has the theoretical prediction (see \cite{my},
e.g., for the full formula)
$$
\Delta M_s^{\rm SM} = (14.5\pm 6.3) {\rm ps}^{-1}.
$$
Combining this with the experimental bound, one has 
$$
\frac{\Delta M_s^{\rm LR}}{\Delta M_s^{\rm SM}} > \frac{14.3}{14.5 + 2
  \times 6.3} = 0.53.
$$
A pattern of $B_s$ mass and decay width differences like that emerging in the 
SB--LR  would be in favour of experimental studies of the $B_s$ decays
at hadron machines, where small values of $\Delta M_s$ and large values of
$\Delta\Gamma_s$ would be desirable. 

Let us finally illustrate the CP-violating asymmetry
of the decay $B_s\to J/\psi\,\phi$:
\begin{equation}\label{CP-asym}
{\cal A}_{\rm CP}(B_s(t)\to J/\psi\,\phi)\equiv\frac{\Gamma(t)-
\overline{\Gamma}(t)}{\Gamma(t)+\overline{\Gamma}(t)}
=\left[\frac{1-D}{F_+(t)+D F_-(t)}\right]
\sin(\Delta M_s t)\,\sin(\arg M_{12}^{B_s}),
\end{equation}
where $\Gamma(t)$ and $\overline{\Gamma}(t)$
denote the time-dependent rates for decays of initially, i.e.\ at $t=0$,
present $B^0_s$ and $\overline{B^0_s}$ mesons into $J/\psi\,\phi$ final
states, respectively. The remaining quantities are defined as
\begin{equation}
D\equiv\frac{|A_{\perp}(0)|^2}{|A_0(0)|^2 + |A_{\|}(0)|^2}\,,
\end{equation}
and
\begin{equation}
F_{\pm}(t)\equiv\frac{1}{2}\left[\left(1\pm\cos(\arg M_{12}^{B_s})\right)
e^{+\Delta\Gamma_s t/2}+\left(1\mp\cos(\arg M_{12}^{B_s})\right)
e^{-\Delta\Gamma_s t/2}\right].
\end{equation}
Here $A_0(t)$, $A_\perp(t)$ and $A_\parallel(t)$ are linear
polarization amplitudes that describe the CP-even and odd final-state
configurations \cite{rosner}.
Note that we have $F_+(t)=F_-(t)=1$ for a negligible width difference
$\Delta\Gamma_s$. Obviously, the advantage of the ``integrated'' observable
(\ref{CP-asym}) is that it can be measured {\it without} performing an
angular analysis and is thus accessible at HERA-B and Tevatron Run
II. The disadvantage is of course that  it
also depends on the hadronic quantity
$D$, which precludes a theoretically clean extraction of $\arg
M_{12}^{B_s}$ from
(\ref{CP-asym}). However, this feature does not limit the power of this
CP asymmetry to search for indications of new physics, which would be
provided by a  sizeable measured value of (\ref{CP-asym}). Model calculations
of $D$, making use of the factorization hypothesis, typically give
$D=0.1\ldots0.5$ \cite{amol}, which is also in agreement with a recent
analysis of the $B_s\to J/\psi\, \phi$ polarization amplitudes performed
by the CDF collaboration \cite{CDF-schmidt}. A recent calculation of
the relevant hadronic form factors from QCD sum rules on the
light-cone \cite{BB} yields $D=0.33$ in the factorization approximation.
Consequently, the CP-odd
contributions proportional to $|A_{\perp}(0)|^2$ may have a significant
impact on (\ref{CP-asym}). 
In Fig.~3(c), we plot this
CP asymmetry as a function of $t$, for fixed values of $D=0.3$, 
$\sin\arg M_{12}^{B_s} = -0.38$, $\Delta \Gamma_s/\Gamma_s = -0.14$ and 
$\Delta M_s=14.5\,{\rm ps}^{-1}$. Although the $B^0_s$--$\overline{B^0_s}$ 
oscillations are very rapid, as can be seen in this figure, it should be 
possible to resolve them experimentally, for example at the LHC. The first 
extremal value of (\ref{CP-asym}), corresponding to $\Delta M_st =\pi/2$, 
is given to a very good approximation by
\begin{equation}\label{ACP-def}
A_{\rm CP}(B_s\to J/\psi\,\phi)=\left(\frac{1-D}{1+D}\right)\sin(
\arg M_{12}^{B_s}),
\end{equation}
which would also fix the magnitude of the $B_s\to J/\psi\,\phi$ CP
asymmetry (\ref{CP-asym}) in the case of a negligible width difference 
$\Delta\Gamma_s$. In Fig.~3(d), we show the prediction of the 
SB--LR  for (\ref{ACP-def}) as a function of the hadronic parameter 
$D$. For a value of $D=0.3$, the CP asymmetry may be as large as 
$-25\%$. The dilution through the hadronic parameter $D$ is not effective
in the case of the CP-violating observables of the $B_s\to J/\psi[\to l^+l^-]
\,\phi[\to K^+K^-]$ angular distribution, which allow one to probe
$\sin(\arg M_{12}^{B_s})$ directly. I thus quote as a second specific
and testable prediction of the SB--LR:
$$
\fbox{$a_{\rm CP}^{\rm SB-LR}(B_s\to J/\psi\phi) \approx -(10-40)\%
  \Longleftrightarrow a_{\rm CP}^{\rm SM}(B_s\to J/\psi\phi)\sim 10^{-2}$}
$$

\section{Summary}

In this talk I have presented 
a detailed investigation of the present status of the
left--right symmetrical model with spontaneous CP violation, based on the
gauge group \model. The parameter space of this model includes the
masses of the predominantly right-handed charged gauge boson, $M_2$,
and of FC neutral and charged Higgs bosons, which we have assumed to be 
degenerate with a common mass $M_H$. Also included are the parameter $\beta$,
which measures the size of the VEV of the Higgs bidoublet $\Phi$ that 
characterizes the spontaneous breakdown of CP symmetry, and the
64-fold discrete ambiguity of the CKM phases due to different quark mass
signatures. In contrast to previous publications, in which the constraints on
the model from K and B physics were treated separately, our paper
\cite{my} is the first one to consider them in a coherent way and to
use the exact results for the CKM phases instead of the small phase 
approximation. We have concentrated on experimental constraints
imposed by the mass differences $\Delta
M_{K,B}$ and observables describing CP violation, i.e.\ $\epsilon_K$,
$\epsilon'/\epsilon$ and $a_{\rm CP}(B_d\to J/\psi K_S)$. In view of the large
theoretical uncertainties, we only use the sign, but not the absolute value of
Re$\,(\epsilon'/\epsilon)$ as a constraint, and we do not use the
electric dipole moment of the neutron. Our main finding is
that, although the K and B constraints can be met {\em separately} by
a large range of input parameters, it is their {\em combination} that
severely restricts the model. We find in particular
that the CP violating observables $\epsilon_K$ and $a_{\rm CP}(B_d\to 
J/\psi K_S)$ are crucial: the sets of input parameters that pass
the constraints imposed by the meson mass differences $\Delta M_{K,B}$
yield to a large majority {\em opposite signs} of $\epsilon_K$ and
 $a_{\rm CP}(B_d\to J/\psi K_S)$. 

We have also performed an analysis of mixing-induced CP-violating effects in 
$B_s\to D_s^+D^-_s$, $J/\psi\, \eta^{(')}$, $J/\psi\, \phi$ decays
and have demonstrated that 
the corresponding CP asymmetries may be as large as ${\cal O}(40\%)$, 
whereas the SM predicts vanishingly small values. Since the 
decay amplitudes of these modes are not significantly affected in the 
SB--LR, direct CP violation remains negligible, as in the SM. 
From an experimental point of view, $B_s\to J/\psi\,\phi$ is a 
particularly promising mode, which is very accessible at B physics 
experiments at hadron machines. We have proposed a simple strategy to 
search for indications of new physics in this transition, which does 
not require an angular analysis of the $J/\psi[\to l^+l^-]$ and 
$\phi[\to K^+K^-]$ decay products. In contrast to the large mixing-induced 
CP asymmetries in the $B_s$ channels, the SB--LR  predicts a small 
value for $a_{\rm CP}(B_d\to J/\psi K_{\rm S})$ below 
10\%. Since the $B_s$ decays cannot be explored at the asymmetric 
$e^+$--$e^-$ B factories operating at the $\Upsilon(4S)$ resonance, 
such a pattern would be in favour of hadronic B experiments. We look 
forward to experimental data to check whether this scenario is actually 
realized by Nature. 

We  would like to stress that our study does not claim to be
exhaustive as we did not allow
the most crucial SM input parameters, the CKM angles and quark masses,
to float within their presently allowed ranges. Taking into account
these uncertainties would certainly affect
the phases of the CKM matrices and thus mainly
show up in the CP violating observables, which, as we have shown, are
crucial. It is therefore not to be excluded that an analysis of the
input parameter uncertainties
would result in increasing the viable LR parameter ranges,
but we doubt that it will change the anticorrelation
between the signs of $\epsilon_K$ and $a_{\rm CP}(B_d\to J/\psi K_S)$
which implies a small maximum value of
$a^{\rm SB-LR}_{\rm CP}(B_d\to J/\psi K_S)$  attainable in the model.

\section*{Acknowledgements}

It is a pleasure to thank R. Fleischer, J.-M.\
Fr\`{e}re and J. Matias for enjoyable collaboration on the work
presented here. I also gratefully acknowledge financial support from the
organizers of the Moriond meeting. This work has been supported by DFG 
through a Heisenberg fellowship.

\end{document}